# Spatial symmetry constraint of charge-ordered kagome superconductor CsV$_3$Sb$_5$

Haoxiang Li[1], Yu-Xiao Jiang[2], J. X. Yin[2,*], Sangmoon Yoon[1], Andrew R. Lupini[3], Y. Y. Pai[1], C. S. Nelson[4], A. H. Said[5], Q. W. Yin[6], C. S. Gong[6], Z. J. Tu[6], H. C. Lei[6], Binghai Yan[7], Ziqiang Wang[8], M. Z. Hasan[2], Ho Nyung Lee[1], and H. Miao[1,*]

[1]*Materials Science and Technology Division, Oak Ridge National Laboratory, Oak Ridge, TN 37831, USA*[1]

[2]*Laboratory for Topological Quantum Matter and Advanced Spectroscopy (B7), Department of Physics, Princeton, New Jersey 08544, USA*

[3]*Center for Nanophase Materials Sciences, Oak Ridge National Laboratory, Oak Ridge, TN, USA*

[4]*National Synchrotron Light Source II, Brookhaven National Laboratory, Upton, NY 11973, USA*

[5]*Advanced Photon Source, Argonne National Laboratory, Argonne, Illinois 60439, USA*

[6]*Department of Physics and Beijing Key Laboratory of Opto-Electronic FunctionalMaterials and Micro-devices, Renmin University of China, Beijing, China*

[7]*Department of Condensed Matter Physics, Weizmann Institute of Science, Rehovot 7610001, Israel*

[8]*Department of Physics, Boston College, Chestnut Hill, Massachusetts 02467, USA*

[*]Correspondence should be addressed to miaoh@ornl.gov and jiaxiny@princeton.edu

**Elucidating the symmetry of intertwined orders in exotic superconductors is at the quantum frontier. Recent surface sensitive studies of the topological kagome superconductor CsV$_3$Sb$_5$ discovered a cascade $4a_0$ superlattice below the charge density wave (CDW) ordering temperature, which can be related to the pair density modulations in the superconducting**

---

[1] This manuscript has been authored by UT-Battelle, LLC under Contract No. DE-AC05-00OR22725 with the U.S. Department of Energy. The United States Government retains and the publisher, by accepting the article for publication, acknowledges that the United States Government retains a non-exclusive, paid-up, irrevocable, world-wide license to publish or reproduce the published form of this manuscript, or allow others to do so, for United States Government purposes. The Department of Energy will provide public access to these results of federally sponsored research in accordance with the DOE Public Access Plan (http://energy.gov/downloads/doe-public-access-plan).

state. If the $4a_0$ phase is a bulk and intrinsic property of the kagome lattice, this would form a striking analogy to the stripe order and pair density wave discovered in the cuprate high-temperature superconductors, and the cascade ordering found in twisted bilayer graphene. High-resolution X-ray diffraction has recently been established as an ultra-sensitive probe for bulk translational symmetry-breaking orders, even for short-range orders at the diffusive limit. Here, combining high-resolution X-ray diffraction, scanning tunneling microscopy and scanning transmission electron microscopy, we demonstrate that the $4a_0$ superstructure emerges uniquely on the surface and hence exclude the $4a_0$ phase as the origin of any bulk transport or spectroscopic anomaly. Crucially, we show that our detected 2×2×2 CDW order breaks the bulk rotational symmetry to $C_2$, which can be the driver for the bulk nematic orders and nematic surface superlattices including the $4a_0$ phase. Our high-resolution data impose decisive spatial symmetry constraints on emergent electronic orders in the kagome superconductor $CsV_3Sb_5$.

The kagome lattice, made of corner-shared triangles, has been one of the most exciting topics in quantum physics for seven decades[1]. Initial interests in the kagome lattice were focused on geometrical spin frustrations to realize quantum spin liquids[2]. More recent studies show that the electronic structure of the kagome lattice is also highly nontrivial, resulting in a rich interplay between topology, geometry, and correlations[3]. A particular focus is for kagome materials near the van Hove filling, where the high density of states together with the orthogonal wave-functions on sub-lattices are proposed to support unconventional symmetry breaking orders such as the $p$-wave charge/spin density waves, $d$-wave Pomeranchuk instability, and $f$-wave superconductivity[4-9].

Recently, a topological kagome metal $A$V$_3$Sb$_5$ ($A$=K, Rb, Cs), possessing a vanadium kagome layer (Fig. 1**a**), was discovered to host intertwined electronic orders. Figures 1**b-d** depict the temperature dependent symmetry-breaking of CsV$_3$Sb$_5$. At room temperature, CsV$_3$Sb$_5$ has an ideal kagome structure that has $D_{6h}$ space group and preserves the time-reversal symmetry, $\mathcal{T}$. Below $T_{\text{CDW}} \sim 92$ K, a three-dimensional charge density wave[10-16] accompanied with an (inverse) star-of-David lattice distortion emerges[9,11,17,18]. Near the same temperature, evidence of $\mathcal{T}$ breaking has been reported in both surface and bulk sensitive probes[11,13,15,19-22], supporting the existence of a CDW with imaginary hopping terms[7-9]. At $T_{\text{C}} \sim 3$ K, CsV$_3$Sb$_5$ enters a superconducting phase that breaks the $U(1)$ gauge symmetry. Interestingly, at an intermediate temperature, $T^* \sim 60$ K, an additional unidirectional 1×4 superlattice is observed by scanning tunneling spectroscopy studies[23-25]. In these surface-sensitive studies, the static 1×4 phase intertwines with CDW and superconducting orders, and thus has been proposed as an analogy to the stripe order and pair density wave in cuprates[26,27], as well as cascade ordering in twisted bilayer graphene[28]. In addition, the 1×4 phase has also been proposed to be associated with the bulk electronic nematicity[23,24]. An outstanding question is that whether the 1×4 phase is a bulk phenomenon, and thus an intrinsic property of each kagome lattice layer.

As shown in Fig. 1**c** and **d**, the 1×4 phase is transparent to thermodynamic measurement. However, anomalies below $T^*$ have been observed in Raman scattering[10,29], ultrafast pump-probe reflectivity[16,25], muon spin relaxation[21], transport[30], and nuclear magnetic/quadruple resonance (NMR/NQR)[31] measurements, pointing to potential connections between the 1×4 phase and the $T^*$ anomaly. In addition, while the three-dimensional 2×2×2 CDW has been identified below $T_{\text{CDW}}$[10,12,15,18], more recent hard x-ray diffraction found evidence of a 2×2×4 superstructure at low temperature[14], adding another layer of complexity to the interplay between electronic symmetry breaking and lattice distortions. Here we combine high precision hard x-ray diffraction, scanning tunneling microscopy and scanning transmission electron microscopy to elucidate the nature of these symmetry breaking orders and discuss their relationship with bulk anomalies.

We first establish the surface modulations of $CsV_3Sb_5$ with scanning tunneling microscopy at 4.2 K. Figure 1**e** shows the topographic image of the clean Sb surface. In agreement with previous studies[23-25], the Fourier transformation of the topography, shown in Fig. 1**f**, reveals both 2×2 and 1×4 superlattice modulations. Figure 1**g** and **h** show d$I$/d$V$ images at ±20 meV, respectively. As highlighted between the black lines, these two spectroscopic maps reveal stripe-like features with an intensity reversal, and 1×4 stripe charge modulations. Therefore, both the topographic imaging and spectroscopic imaging confirms the existence of the 1×4 superlattice.

To search for the superstructure peaks in the bulk crystal of $CsV_3Sb_5$, we perform high-precision hard x-ray diffraction in transmission geometry as shown in Fig. 2**a**. The bandwidth of incident energy is 1 meV; this is done using a high-resolution monochromator (HRM) that consists of six silicon crystals[32,33]. The meV-beam is then focused to a lateral size of 15×35 μm$^2$ (vertical×horizontal). A total $\Delta E$ =1.5 meV energy resolution is achieved by matching the incident photon energy, $h\nu$=23.72 keV, to the backscattering energy of Si(12,12,12). The extremely high energy resolution filters out background intensity[34] and false peaks appearing in conventional X-ray diffraction (see Extended Data Fig. 1), allowing experimental detection of static translational symmetry breaking even in the diffusive limit[35-37]. Figure 2**b** shows high-precision hard x-ray diffraction measurement of stripe ordered $La_{1.875}Ba_{0.125}CuO_4$ and optimally doped $La_{1.83}Sr_{0.17}CuO_4$, where a 4$a_0$ CDW correlation length less than 20 Å  has been successfully detected[35-40]. We therefore utilize this ultra-high-resolution technology to characterize the kagome superconductor $CsV_3Sb_5$. Figure 2**c** shows the scattering trajectories along the high symmetry directions, *i.e.* [1,0,0] (*H*-scan), [0,1,0] (*K*-scan), [1, -1,0] and [1,1,0] (*HK*-scan) in reciprocal lattice unit (r.l.u.). Figure 2**d** present base temperature (*T*=10 K) *H*, *K* and *HK*-scans at the *L*=0 and *L*=2 plane. Peaks at *H*=0.5 and/or *K*=0.5 are corresponding to the 2×2×2 CDW superlattice peaks. Through all scattering trajectories, we do not detect any 1×4 superstructure peak located at *q*=0.25 or 0.75 r.l.u. This observation is in stark contrast with the STM result shown in Fig. 1**f**, where the 1×4 superstructure peak intensity is comparable with the in-plane 2×2 CDW peak in topographic imaging and is even stronger than the in-plane 2×2 CDW peaks in spectroscopic imaging. In addition, STM also fails to detect the 1×4 in the Cs terminated surface, where the in-plane 2×2 CDW peaks clearly exist[23-25]. These results taken together demonstrate that the 1×4 phase in $CsV_3Sb_5$ is a surface/interface effect that is mainly associated with the Sb terminated surface.

The surface origin of the 1×4 phase is consistent with a recent density functional theory study, where the 1×4 superstructure is stabilized by the relative shift between V-atoms and surface Sb-atoms[25]. The absence of the 1×4 phase in bulk $CsV_3Sb_5$ raises questions on the origin of $T^*$ anomalies in other bulk sensitive measurements[10,16,21,25,30,31]. Recent NMR/NQR studies[31] suggest that the $T^*$ anomaly could be related to the 2×2×4 superstructure observed in a previous XRD study[14]. To further explore this possibility, we perform the XRD measurement with a reflection geometry to search for superlattice peaks along the out-of-plane direction. Interestingly, as shown in Fig. 3**a**, while we confirm the *L*=0.25 superlattice peaks, these peaks persist up to room temperature in our high precision measurement. In contrast, the CDW peak at *L*=0.5 disappears above $T_{CDW}$, consistent with transport and specific heat measurement shown in Fig. 1**c** and **d**. We thus determine that the 2×2×4 superstructure is also irrelevant to the $T^*$ anomaly and likely a consequence of $4c_0$ stacking fault in this material ($c_0$ is the out-of-plane lattice constant). We further find that the correlation length, defined as the inverse half-width-at-half-maximum (1/HWHM)[10,35,41], of $L$=0.25 superlattice peak is 257 Å, on a similar scale to the CDW correlation length of 480 Å ($L$=0.5), suggesting mesoscopic structural phase separation. Our scanning transmission electron microscope measurement at room temperature shows inhomogeneous spacing between adjacent kagome layers in the cross-sectional view (Fig. 3**b**). The variation of lattice modulation is over 8%, suggesting $c$-axis defects in $CsV_3Sb_5$ (see Supplementary Information S3).

The high precision XRD study excludes translational symmetry breaking as the origin of $T^*$ anomalies. It is important to note that while the in-plane 2×2 charge modulation preserves $C_6$ symmetry (Fig. 3**d**), the 3-dimensional 2×2×2 superstructure breaks the rotational symmetry down to $C_2$ as depicted in Fig. 3**c-e** [17], where the red and blue arrows correspond to anti-phase lattice distortions[16,18]. This twofold rotational symmetry is a consequence of the $\pi$ phase shift between the adjacent kagome layers[9,17], and naturally provide a unidirectional crystal field below $T_{CDW}$, which can be a driver for electronic nematicity. We therefore conclude that neither transnational symmetry nor rotational symmetry breaking directly drives the $T^*$ anomaly. However, the rotational symmetry breaking in the bulk CDW may facilitate the formation of the nematic order on the surface, which potentially results in the 1×4 surface superlattice. On the other hand, recent evidence of time-reversal symmetry breaking has been observed in both $KV_3Sb_5$ and $CsV_3Sb_5$[11,13,15,19-21], meaning it will be interesting to evaluate the interplay between time-reversal symmetry and spatial symmetry and their relevance to the $T^*$ anomaly. In short, the decisive high-precision XRD data together with STM and TEM measurements point to the surface nature of the 1×4 phase and suggest that the 2×2×2 CDW featuring intrinsic nematicity is the only spatial symmetry breaking below the room temperature in bulk kagome superconductor $CsV_3Sb_5$.

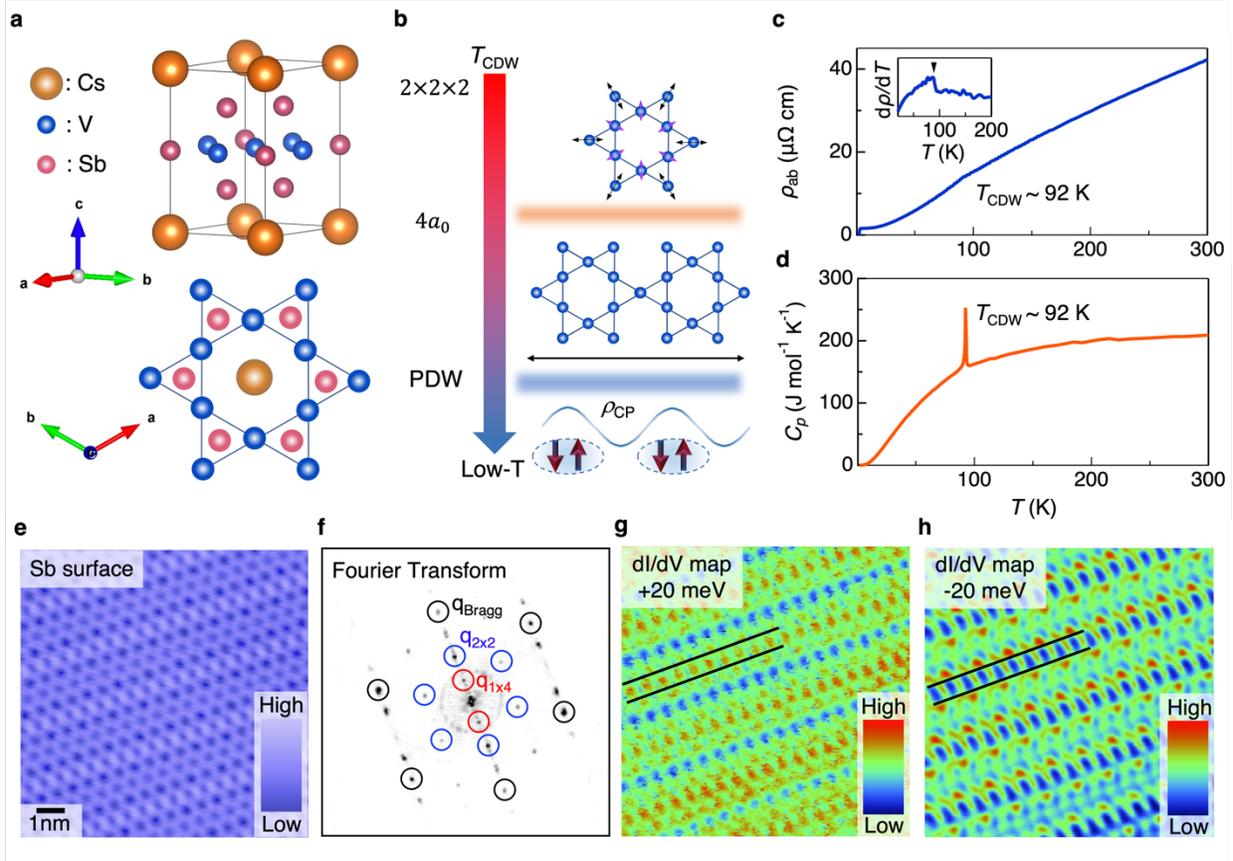

**Fig. 1. The crystal structure and temperature dependent symmetry breaking phases of $CsV_3Sb_5$. a.** the crystal structure of $CsV_3Sb_5$ (space group P6/mmm, No. 191). It consists of V-Sb slabs that are separated by Cs. The V-Sb slab contains a V-kagome lattice and two Sb sites laying in the kagome plane and above the V-triangle. **b.** Schematic of temperature-dependent symmetry-breaking phases in $CsV_3Sb_5$. **c** and **d** show the resistivity and specific heat of $CsV_3Sb_5$, respectively. The inset in panel **c** shows the derivative $d\rho/dT$ curve where the black triangle marks the CDW transition. **e.** Topographic image of a clean Sb surface (V = -100 mV, I = 0.5 nA). **f.** Fourier transform of the topography showing Bragg peaks and charge ordering vector peaks. The in-plane

2×2 CDW, 1×4 charge order and Bragg peaks are highlighted by blue and red circles, respectively. **g** and **h** show atomically resolved d$I$/d$V$ images for the same Sb surface at energy E=-20 meV and E=+20 meV, respectively.

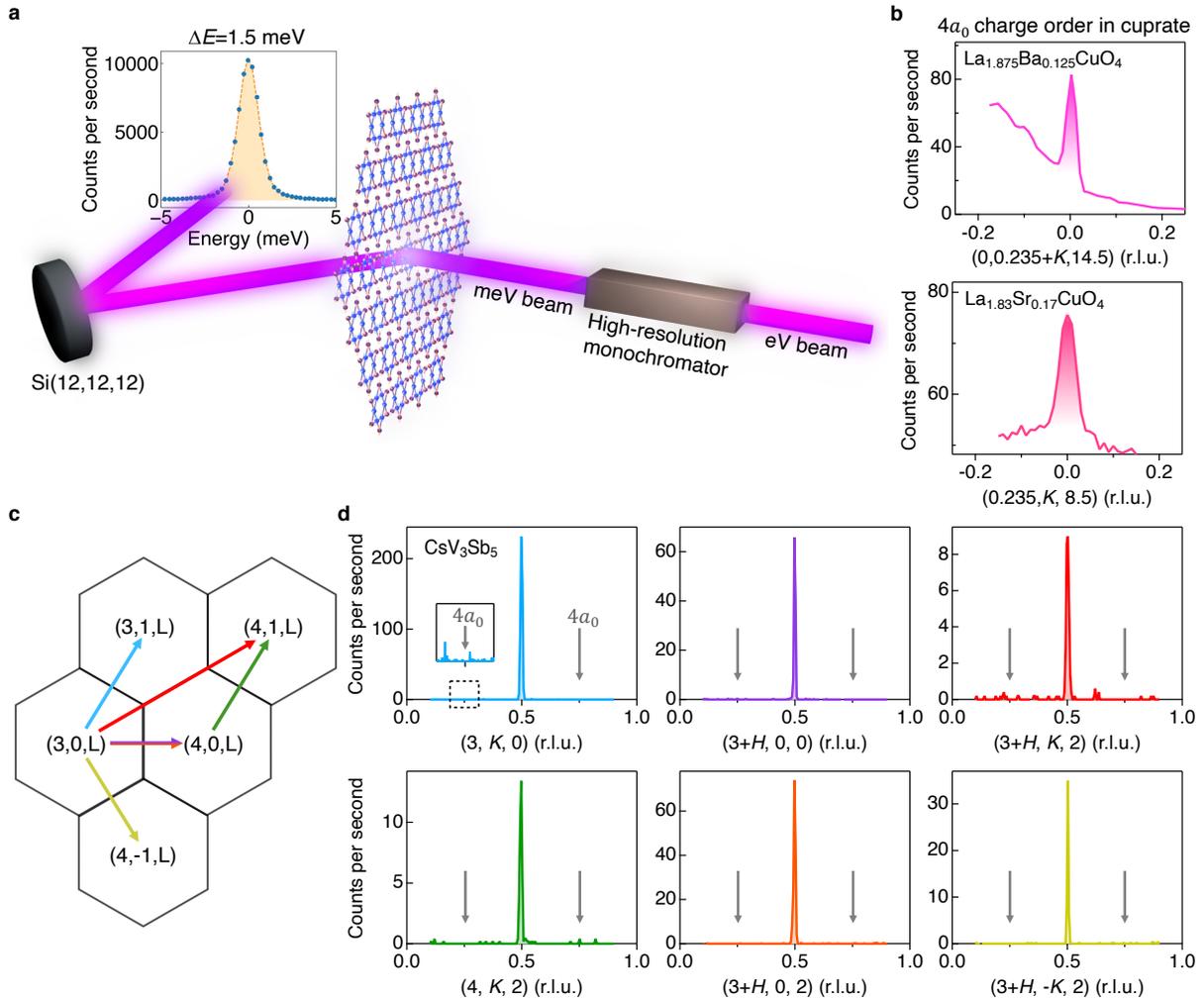

**Fig. 2. Absence of 1×4 superlattice peak in high precision hard x-ray diffraction of CsV₃Sb₅.** **a**. Schematic of the experimental setup for the hard x-ray diffraction measurement with meV resolution. The inset spectrum reveals the total energy resolution of ~1.5 meV (FWHM) (see Supplementary Information S2 for more details). **b**. Extracted high precision hard x-ray diffraction data of the charge ordered La$_{1.875}$Ba$_{0.125}$CuO$_4$ [35] and La$_{1.83}$Sr$_{0.17}$CuO$_4$ [36]. **c**. Schematic indicating the scattering trajectories for scans shown in panel **d**. Hard x-ray diffraction measurements of CsV₃Sb₅ shown in panel **d** were performed at 10 K and covered all in-plane high symmetry directions with zero and finite $L$ components. The 2×2×2 CDW peaks can be clearly resolved in all scans, but there is no peak feature at $q$=0.25 or 0.75 (highlighted by the grey arrows), demonstrating the absence of the 1×4 superlattice in bulk CsV$_3$Sb$_5$.

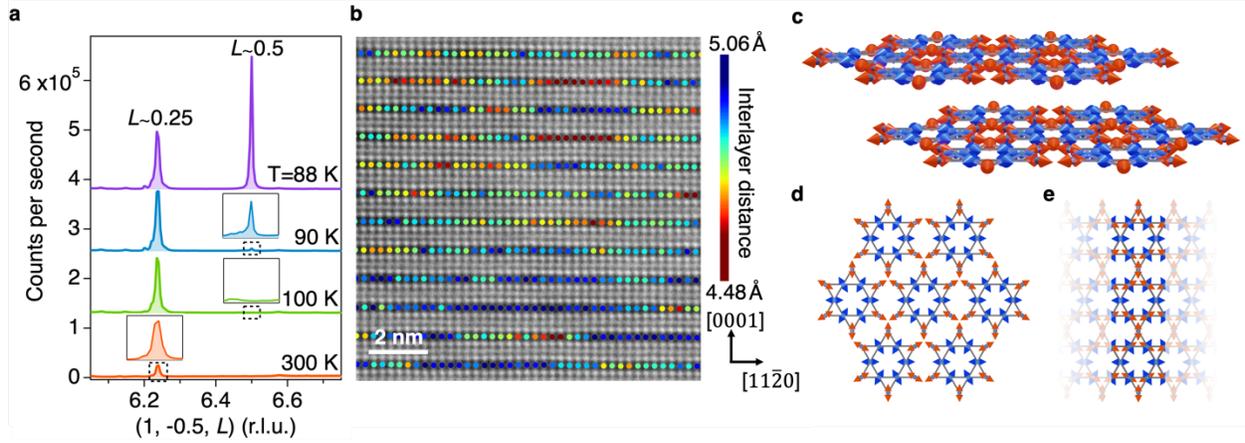

**Fig. 3. Out-of-plane superstructure peaks and Unidirectional crystal field below $T_{CDW}$. a.** Temperature dependent XRD measurement from **Q**=(1, -0.5, 6.05) to (1, -0.5, 6.75). The CDW peak at $L$=6.5 disappear above the $T_{CDW}$~92 K, whereas the superstructure peak at $L$=6.25 persist up to room temperature. By fitting the superstructure peaks with a Lorenzian-squared function[35,41], we extract the HWHM of each peak, which is 0.006, 0.003 r.l.u. for $L$=0.25, 0.5 r.l.u., respectively. The correlation length, defined as inverse HWHM (1/HWHM), is 257, 480 Å for $L$=0.25, 0.5 r.l.u., respectively. **b.** Scanning transmission electron microscopy image in the cross-section geometries with a color code map of the interlayer distance between V$_3$Sb$_5$ layers, where the distances are represented on the Cs atomic columns with a color code. The lattice of CsV$_3$Sb$_5$ is significantly modulated along the out-of-plane axis while it remains highly periodic in the in-plane directions. **c.** A 3-dimensional view of the 2×2×2 CDW with star-of-David (SD)/inverse-star-of-David (ISD) lattice distortions on the V kagome layer (panel **d**) and the out-of-plane π phase shift between the two V kagome layers[16-18]. The red and blue arrows represent out-of-phase atomic distortions moving toward/outward the center of the V-hexagon, respectively. **e.** Top-down view of the kagome sublattices with the interlayer π phase shift. The $C_6$ symmetry is preserved in the single kagome layer (panel **d**) but breaks down to $C_2$ due to the phase shift between kagome layers.

**Method**

**Sample preparation and characterizations**

Single crystals of $CsV_3Sb_5$ were grown from Cs ingot (purity 99.9 %), V powder (purity 99.9 %) and Sb grains (purity 99.999 %) using the self-flux method[42]. The mixture was put into an alumina crucible and sealed in a quartz ampoule under partial argon atmosphere. The sealed quartz ampoule was heated to 1273 K for 12 h and soaked there for 24 h. Then it was cooled down to 1173 K at 50 K/h and further to 923 K at a slowly rate. Finally, the ampoule was taken out from the furnace and decanted with a centrifuge to separate $CsV_3Sb_5$ single crystals from the flux. The obtained crystals have a typical size of $2\times2\times0.02$ mm$^3$. CDW transition is clearly observed in transport and specific heat measurement shown in Fig. 1**c** and **d**, respectively.

**Hard X-ray diffraction**

High-precision X-ray scattering measurements were performed at 30-ID-C (HERIX) of Advanced Photon Source (APS) and 4-ID beamline of NSLS-II. Data shown in Fig. 2**d** was taken at 30-ID-C (HERIX), where the highly monochromatic x-ray beam of incident energy $E_i$ = 23.7 keV (l= 0.5226 Å) was focused on the sample with a beam cross section of $35\times15\mu m^2$ (horizontal × vertical). The total energy resolution of the monochromatic x-ray beam and analyzer crystals was $\Delta E\sim1.5$ meV (full width at half maximum). The measurements were performed in transmission geometry. Data shown in Fig. 3**a** was taken at 4-ID beamline of NSLS-II. The photon energy, which was selected by a cryogenically-cooled Si(111) double-crystal monochromator, was 11.47~keV. The sample was mounted in a closed-cycle displex cryostat in a vertical reflection scattering geometry, and the sigma-sigma scattering channel was measured using a MgO(440) polarization analyzer and silicon drift detector. Due to the reflection geometry and lower incident photon energy, the attenuation length for data shown in Fig. 3**a**, is about 10 μm, a factor of 10 smaller than the data shown in Fig. 2**d**.

**Scanning tunnelling microscopy**

Single crystals with size up to 2 mm × 2 mm were cleaved mechanically in situ at 77 K in ultra-high vacuum conditions, and then immediately inserted into the microscope head, already at Helium-4 base temperature (4.2 K). Topographic images in this work were taken with the tunnelling junction set up $V$=100 mV and $I$=0.05 nA for exploration of areas typically 400 nm × 400 nm. When we found atomically flat and defect-free areas, we took topographic images with the tunnelling junction set up $V$=100 mV and $I$=0.5 nA to resolve the atomic lattice structure as demonstrated in the paper. Tunnelling conductance spectra were obtained with an Ir/Pt tip using standard lock-in amplifier techniques with a lock-in frequency of 997 Hz and a junction set up of $V$=50 mV and $I$=0.5 nA, and a root-mean-square oscillation voltage of 0.3 mV. Tunnelling conductance maps were obtained with a junction set up of $V$=50 mV and $I$=0.3 nA, and a root-mean-square oscillation voltage of 5 mV.

**Scanning transmission electron microscopy**

Scanning transmission electron microscopy (STEM) measurements were performed on a Nion UltraSTEM200 operating at 200 kV. The microscope is equipped with a cold field emission gun and a corrector of third- and fifth-order aberrations for sub Ångstrom resolution. An electron probe with a convergence half-angle of 30 mrad was used, and the inner angle of the high angle annular dark-field (HAADF) STEM was approximately 65 mrad. Plan-view and cross-section STEM specimens were prepared by drop-casting followed by sonication, and by focused ion beam followed by Ar ion milling, respectively. HAADF STEM images were simulated using the multislice-based QSTEM code. The interlayer distances between $V_3Sb_5$ layers were estimated using a homemade python script from the cross-sectional-view HAADF STEM image. The steps of the post-processing procedures were: the coordinates of all atomic columns were measured using the blob-detection function in the skimage library, the Sb atomic columns were sorted based on the scattering intensities, and the interlayer distances were calculated from the measured atomic coordinates and represented on the Cs atomic columns with a color code.

**Data availability:** The data that support the findings of this study are available from the corresponding author on reasonable request.

**Acknowledgement:** We thank Kun Jiang, Jiaqiang Yan and Michael McGuire for stimulating discussions. This research was sponsored by the U.S. Department of Energy, Office of Science, Basic Energy Sciences, Materials Sciences and Engineering Division (x-ray diffraction). H.C.L. was supported by National Natural Science Foundation of China (Grant No. 11822412 and 11774423), Ministry of Science and Technology of China (Grant No. 2018YFE0202600 and 2016YFA0300504) and Beijing Natural Science Foundation (Grant No. Z200005). Z.Q.W is supported by the U.S. Department of Energy, Basic Energy Sciences Grant No. DE-FG02-

99ER45747 (sample growth). B.Y. acknowledges the financial support by the European Research Council (ERC Consolidator Grant, No. 815869) and the Israel Science Foundation (ISF No. 1251/19 and No. 2932/21) (Theory). This research uses resources (meV-IXS experiment at beam line 30-ID) of the Advanced Photon Source, a U.S. DOE Office of Science User Facility operated for the DOE Office of Science by Argonne National Laboratory under Contract No. DE-AC02-06CH11357. Hard-x-ray diffraction measurements use resources at the 4-ID beam line of the National Synchrotron Light Source II, a U.S. Department of Energy Office of Science User Facility operated for the DOE Office of Science by Brookhaven National Laboratory under Contract No. DE-SC0012704. Extraordinary facility operations are supported in part by the DOE Office of Science through the National Virtual Biotechnology Laboratory, a consortium of DOE national laboratories focused on the response to COVID-19, with funding provided by the Coronavirus CARES Act.

**Author contributions:** H.L., C.S.N., A.H.S., H. N. L. and H.M. performed the XRD experiment. H.L. and H.M. analyzed the XRD data. Y.X.J., J.X.Y., and M.Z.H. performed STM measurements. S.Y. and A.R.L. conducted the STEM experiment. B.Y. and Z.W. performed the theoretical analysis. Y.M.Y., Q.W.Y., C.S.G., Z.J.T., H.C.L. synthesized the high-quality single crystal samples. H.L., J.X.Y., and H.M. prepared the manuscript with inputs from all authors.

**Competing interests:** The authors declare no competing interests.